# Orbital-dependent modifications of electronic structure across magneto-structural transition in BaFe$_2$As$_2$

T. Shimojima[1], K. Ishizaka[1], Y. Ishida[2,3], N. Katayama[1], K. Ohgushi[1,3], T. Kiss[1], M. Okawa[1], T. Togashi[2], X. -Y. Wang[4], C. -T. Chen[4], S. Watanabe[1], R. Kadota[5], T. Oguchi[5], A. Chainani[2], S. Shin[1,2,3]

[1] *Institute for Solid State Physics (ISSP), University of Tokyo, Kashiwa, Chiba 277-8581, Japan*

[2] *RIKEN SPring-8 Center, Sayo, Sayo, Hyogo 679-5148, Japan*

[3] *JST, TRIP, 5, Sanbancho, Chiyoda, Tokyo 102-0075, Japan*

[4] *Beijing Center for Crystal R&D, Chinese Academy of Science Zhongguancun, Beijing 100190, China*

[5] *Department of Quantum Matter, Graduate School of Advanced Sciences of Matter (ADSM), Hiroshima University, 1-3-1 Kagamiyama, Higashi-Hiroshima 739-8530, Japan*

*Abstract,*

Laser angle-resolved photoemission spectroscopy (ARPES) is employed to investigate the temperature ($T$) dependence of the electronic structure in BaFe$_2$As$_2$ across the magneto-structural transition at $T_N \sim 140$ K. A drastic transformation in Fermi surface (FS) shape across $T_N$ is observed, as expected by first-principles band calculations. Polarization-dependent ARPES and band calculations consistently indicate that the observed FSs at k$_z \sim \pi$ in the low-$T$ antiferromagnetic (AF) state are dominated by the Fe3d$_{zx}$ orbital, leading to the two-fold electronic structure. These results indicate that magneto-structural transition in BaFe$_2$As$_2$ accompanies orbital-dependent modifications in the electronic structure.

The iron-pnictide superconductors (*1*) exhibit high transition temperature ($T_c$) of up to 55 K (*2*) thus casting doubt on the relevance of phonon-mediated Bardeen-Cooper-Schrieffer pairing mechanism (*3*). The iron-pnictides are the second class of compounds after the copper oxides, which are considered as candidate materials for room temperature superconductors. Superconductivity in iron-pnictides emerges via application of hydrostatic pressure (*4*) or carrier doping (*5*) into the parent materials which undergo a structural transition, where a high-*T* paramagnetic (PM) metal becomes a low-*T* AF metal (*6*). While antiferromagnetism of copper-oxide superconductors is derived from a localized Mott-insulator, its role for superconductivity in the iron-pnictides is intriguing (*7*), particularly because the high-*T* undoped phase is an itinerant metal (*8-10*). The possible roles of multi-band character in terms of the five Fe 3d orbitals forming multiple FSs should be also considered seriously, in contrast to single-band copper oxides. Some theoretical studies proposed that the AF spin fluctuation originating from the nesting between disconnected FSs plays an important role in the superconducting mechanism (*11,12*). Hence, the elucidation of the electronic structure of the parent materials, especially in the AF low-*T* phase, is highly desired in order to unravel the unconventional superconductivity in the iron-pnictides. Several ARPES studies have been performed across the magneto-structural transition. (*13-15*). In these reports, the *T*-dependence of the electronic structure is understood by the exchange splitting (*13,14*) or anisotropic spin density wave gap opening (15) which appear in AF phase. However, it is difficult to discuss the impact of the magneto-structural transition on the electronic structure of the parent compounds due to the mixing of twinned domains in the orthorhombic structure (*16-18*). Since the typical domain size is very small (~ 1 μm), the electronic structures observed by macroscopic probes will inevitably overlap with those rotated by 90-degrees.

In this work, we report the T-dependence of the electronic structure of an iron-pnictide superconductor parent material BaFe$_2$As$_2$ by employing linearly polarized vacuum-ultra-violet (VUV) laser as a photon source for ARPES (*19-21*). While the size of laser spot (~ 200 μm) is much larger than that of orthorhombic domains below $T_N$, we could separately extract the electronic structure from each of the twinned domains, owing to matrix element effects which make ARPES intensity strongly dependent on light polarization and sample geometry. As explained later, all the data appearing in this Letter simply reflect the electronic structure from single domains. We found a drastic transformation in FS shape across $T_N$ which is qualitatively explained by the first-principles local density approximation (LDA) band structure calculation. Polarization-dependent ARPES and LDA calculations consistently indicate that the observed cut of the FSs in AF state are dominated by $Fe3d_{zx}$ orbital symmetry. These results highlight that $BaFe_2As_2$ shows strongly orbital-dependent modifications of the electronic structure across the magneto-structural transitions, and is consistent with recent theory showing orbital ordering in the pnictides(*22*).

Single crystals of $BaFe_2As_2$ were grown in a $Al_2O_3$ crucible using the self-flux method, as described in Ref. 16. $BaFe_2As_2$ exhibits stripe-type AF ordering of Fe spins below $T_N \sim 140$ K, accompanied by a structural phase transition. The high-$T$ PM metal tetragonal structure has lattice parameters of $a^T = b^T = 3.9625$ Å and $c^T = 13.0168$ Å, while the stripe-type AF ordered orthorhombic structure for $T < T_N$ has $a^o = 5.6146$ Å, $b^o = 5.5742$ Å, $c^o = 12.9453$ Å (*23*), as shown in Fig. 1(a) and (b). Neutron diffraction studies have shown that the magnetic modulation vector is $Q_m = (101)_o$ in the above orthorhombic setting, with the spin orientation parallel to $a^o$ (*6*). Laser ARPES measurements were performed on a spectrometer built using a VG-Scienta R4000WAL electron analyzer and a VUV-laser of hν = 6.994 eV as a photon source (*21*). Spot size of the VUV-laser is 200 - 300 μm. Using the λ/2 (half-wave) plate, we can rotate the light polarization vector and obtain *s*- or *p*-polarized light without changing the optical path (see Fig. 2). Unless noted, the measurements were performed in p-polarized configuration. The energy resolution was set to 10 meV to get high count rate. The spectra were reproducible over measurement cycles of 6 hours. Fermi level ($E_F$) of samples was referenced to that of gold film evaporated onto the sample substrate. All measurements were done on surfaces obtained by cleaving samples at 180 K in an ultra high vacuum of better than $2 \times 10^{-11}$ Torr.

In Figure 1(c) and (d), we show the FSs measured by p-polarized laser above and below $T_N$, respectively. Above $T_N$, the observed FS shows a nearly circular shape with four-fold symmetry. Its shape and size are very consistent with previous ARPES studies (*9,10,13-15,24*). In contrast, on lowering temperature below $T_N$, the FS gets drastically split into a relatively large pair of FSs along $a^{o*}$ and a pair of small FSs along $b^{o*}$, thus resulting in significantly modified FSs with two-fold symmetry. Furthermore, when the linear polarization is changed from p to s at the same sample position, the two-fold FSs in the low-*T* AF state rotate by 90-degrees (Fig. 1 (d) and (f)), while very little change is observed above $T_N$ (Fig. 1 (c) and (e)).

Before comparing our ARPES results with LDA calculations, we show that all data at $T < T_N$ presented in this Letter are separately observed electronic structures of respective single domains (domain A and B) in the twinned structure (Fig.2 (a)). Considering the selection rules (*25,26*), the complete 90-degrees rotation of the FSs implies that FS of each single domain has particular orbital character and is alternatively active to s- or p-polarization. In our experimental configuration, with the incidence plane corresponding to the mirror planes of the crystal, (Fig. 2 (b) and (c)), the initial states only having even (odd) parity with respect to the mirror plane are active to *p*(*s*) – polarization. (*25,26*) The detectable d-orbitals following this selection rule are thus summarized for each possible relation between the polarization and orthorhombic orientation, as shown in Fig.2 (b) and (c). Here we note that orthorhombic orientation in domain A is 90-degrees rotated with respect to domain B, and that the permissible combination of experimental geometries is limited to 'case 1' or 'case 2'. Thus, we can observe the 90-degrees rotation of FSs, only if the FS is composed of single $d_{zx}$ (case 1) or $d_{yz}$ (case 2) orbital (*x*, *y*, and *z* are coordinates along the crystal axes of the orthorhombic setting $a^o$, $b^o$, and $c^o$, respectively). The polarization-dependent ARPES results consistently indicate the independent observation of the FS from each single domain. The present results are consistent with recent theoretical studies (*22*) which show the importance of orbital ordering in the pnictides. It causes the degeneracy between $d_{zx}$ and $d_{yz}$ orbitals to get broken by the tetragonal to orthorhombic transition, and leads to a difference in occupancy of the $d_{zx}$ and $d_{yz}$ derived bands, which is important for the superexchange terms. (*22*)

Now we show in figure 1 (g) and (h) the FSs obtained from LDA calculation below and above $T_N$ using experimental lattice constant, (*23*) respectively. The stripe-type AF ordering of Fe spins is taken into account in the calculation below $T_N$. (*27*) As compared to the fairly 2-dimensional FS in the high-*T* PM state, the low-*T* phase FSs are 3-dimensional with large anisotropy. It is consistent with the recent quantum oscillation measurements at low-*T* indicates the existence of highly 3-dimensional FSs. (*29,30*) If we choose the cut centered at Z point (Z-plane in Fig. 1(g) and (h), respectively), the transformation in the FS shape across the magneto-structural transition revealed by ARPES is fairly reproduced by LDA calculation. This is also justified by comparing with recent 3-D (hν-dependent) ARPES studies (31,32) which reports electron-like dispersion corresponding to the V-shape dispersion in our data (Fig. 3(a), discussed later), and which is observed only near $k_z = \pi$. The obtained FSs in Fig. 1 are thus consistently explained as the result of probing the Z-plane in the high and low-*T* phases using laser ARPES with hν = 6.994 eV. This correspondence confirms that the observed polarization-dependent FS is indeed reflecting the electronic structure from single domain. We have also confirmed that LDA calculation for a low-$T$ orthorhombic structure without stripe type AF order induces almost negligible change in FS shape (not shown) compared to the high-$T$ phase. This indicates that the FS transformation is mainly induced by the spin-stripe AF order.

We next show the band dispersions along high symmetry lines. Fig. 3(a) and (b) show the ARPES intensity images of BaFe$_2$As$_2$ at 30 K along Z-Γ' and Z-T directions (Γ' point corresponds to Γ point in the second Brillouin zone (BZ) : see supplementary information). Peak positions of momentum distribution curves and energy distribution curves are superimposed on the intensity images by black and gray circles, respectively. By tracing the peak positions, a V-shaped (one hole and one electron) dispersion can be recognized near $E_F$. Such characteristic dispersions are consistent with the LDA calculation in AF state as indicated by blue and red curves in Fig. 3 (c). Furthermore, a hole band sinking below $E_F$ is observed along Z-Γ' direction. We note that the number of the band dispersions near $E_F$ in AF state is better explained by the LDA calculation of the low-$T$ phase rather than a simple folding of the PM band structure. If we overlap the ARPES images along both directions taking into account the contribution from both twinned domains, we will have three hole and two electron bands. It shows good agreement with the synchrotron-based ARPES using elliptical polarization reporting at least three hole and one electron bands in AF state (13).

Our LDA calculations show that the Fe 3d$_{zx}$ orbital character dominates the FSs in AF state, consistent with the above-mentioned polarization-dependent ARPES results. Figure 4 shows partial densities of states (PDOS) of Fe 3$d$ and As 4$p$ for tetragonal PM and orthorhombic AF states. To easily compare with each other, x, y, z used in figure 4 are common and correspond to a$^o$, b$^o$, c$^o$, respectively. The high-$T$ PM state exhibits the DOS near $E_F$ consisting of multiple Fe 3$d$ orbitals (Fig.4 (a)). In contrast, FSs in AF state is mainly composed of minority spin component of Fe d$_{zx}$ orbital and other orbital components are pseudogapped near $E_F$ (Fig.4 (b)). The upper panel PDOS originates in minority-spin states, while the lower panel is for majority-spin states. Here we note that the LDA calculation thus supports "case 1" with FSs of d$_{zx}$ character as shown in Figure 2 and we can accordingly determine the orientation of $a^{o*}$ and $b^{o*}$ in Fig. 1 (d) and (f). Even though the As 4$p$ PDOS at $E_F$ in Fig. 4 (c) and (d) is relatively small compared to that of Fe 3$d$ PDOS, one can see that the As 4$p$ PDOS is also reconstructed across $T_N$. In AF state, the As $p_y$ PDOS has negligible intensity near $E_F$ whereas the $p_x$ and $p_z$ PDOSs have finite weight at $E_F$. These results indicate that the FS of BaFe$_2$As$_2$ in AF state is dominated by d$_{zx}$ orbital with finite but small hybridization with As p$_x$ and p$_z$ orbitals, showing that the antiferromagnetism affects all over the FeAs network through multiple d-orbitals. The great difference in the near-$E_F$ DOS between PM and AF state thus brings out the orbital-dependent electronic reconstruction across $T_N$.

Our present result thus concludes that the AF phase of the iron pnictide BaFe$_2$As$_2$ shows the metallic state with clear FSs, in contrast to the cuprates whose AF phase caused by the half-filled single-band Mott transition is an insulator. The strongly orbital-dependent modifications in the electronic structure further distinguishes BaFe$_2$As$_2$ from other conventional AF metals in which such a phenomenon has been rarely discussed. These issues raise the importance of closely investigating the possible role played by each orbital across the transition. It is worth noting that domination of the single orbital component may be also expected in an orbital-selective Mott-transition scenario (*33*), which has been recently discussed for strongly correlated multi-orbital systems. Although the electron correlation in the iron-pnictides is not expected to be as strong as the cuprates (*34,35*), our observation of orbital-poralized two-fold symmetric FSs in the AF phase favors an orbital ordering scenario with strong magnetic anisotropy, which results in the degeneracy between the d$_{zx}$ and d$_{yz}$ orbitals to get broken across T$_N$. (*22*) Further magnetic measurements such as spin-polarized neutron scattering will be necessary for investigating the origin of the magnetic moment in AF state.

In summary, laser-ARPES on BaFe$_2$As$_2$ reveals a drastic transformation in FS shape across $T_N$ which is qualitatively explained by the first-principles band calculations. Polarization-dependent ARPES and band calculations consistently indicate that the observed cut of the FSs in AF state are dominantly composed of the Fe3d$_{zx}$ orbital, in contrast to the multi-orbital character of the high-*T* nonmagnetic phase. We conclude that BaFe$_2$As$_2$ shows strongly orbital-dependent reconstruction of the electronic structure across the magneto-structural transition.

also performed preliminary LDA calculations using a weak staggered magnetic field which results in an artificial reduced magnetic moment of 0.75 muB. We found that the size of FS changes (up to 30 % at $k_z = \pi$) due to the modification of the electronic structure near $E_F$, nevertheless, the shape and number of FS and its dominant orbital component ($d_{zx}$) are unchanged.

**Author Information** Correspondence and requests for materials should be addressed to T.S. (t-shimo@issp.u-tokyo.ac.jp) or S.S. (shin@issp.u-tokyo.ac.jp).

*Figure legends,*

**Figure 1.**

(**a**), Crystal structure of BaFe$_2$As$_2$ in the tetragonal structure (**b**), Crystal and magnetic structure in the stripe-type AF ordered orthorhombic structure. (**c**), (**d**), FS of BaFe$_2$As$_2$ measured by p-polarization at 180 K (above $T_N$) and 30 K (below $T_N$), respectively. (**e**), (**f**), FS of BaFe$_2$As$_2$ measured by s-polarization at 180 K (above $T_N$) and 30 K (below $T_N$), respectively. (**g**), (**h**), Whole FS in the first BZ obtained by LDA calculation considering PM tetragonal and stripe-type AF (*6*) orthorhombic structure, respectively, using the experimentally obtained structural parameters. PM structure is represented by a space group symmetry of I4/mmm, while we use Cccm notation for AF state regarding the Fe sites having opposite spins as nonequivalent.

**Figure 2.**

(**a**), Schematics of orthorhombic twinned structure at $T < T_N$. Orthorhombic orientation in domain A is 90-degrees rotated with respect to domain B. (**b**), (**c**), Two permissible cases of the relation between the polarization and orthorhombic orientation, denoted as "case 1" and "case 2", respectively. Arrows on the square lattice represent the stripe-ordered Fe spins.

**Figure 3.**

(**a**), (**b**), ARPES intensity image along Z-Γ' (a$^o$*) and Z-T (b$^o$*) directions. (**c**), Band structure near $E_F$ obtained from LDA calculation in the stripe-type AF ordered state. Blue and red curves correspond to the V-shaped band dispersions observed in (a) and (b), respectively.

**Figure 4.**

(**a**), (**b**) Calculated Fe 3d PDOS in PM and AF state, respectively. The upper panel PDOS in (b) and (d) originates in minority-spin states, while the lower panel is for majority-spin states. Note that d$_{zx}$ and d$_{yz}$ orbitals are degenerate in the tetragonal PM state and their PDOS are equivalent. (**c**), (**d**), Calculated As p PDOS in PM state and AF state, respectively. p$_x$ and p$_y$ PDOS are equivalent in PM state.

*Figures,*

Figure 1

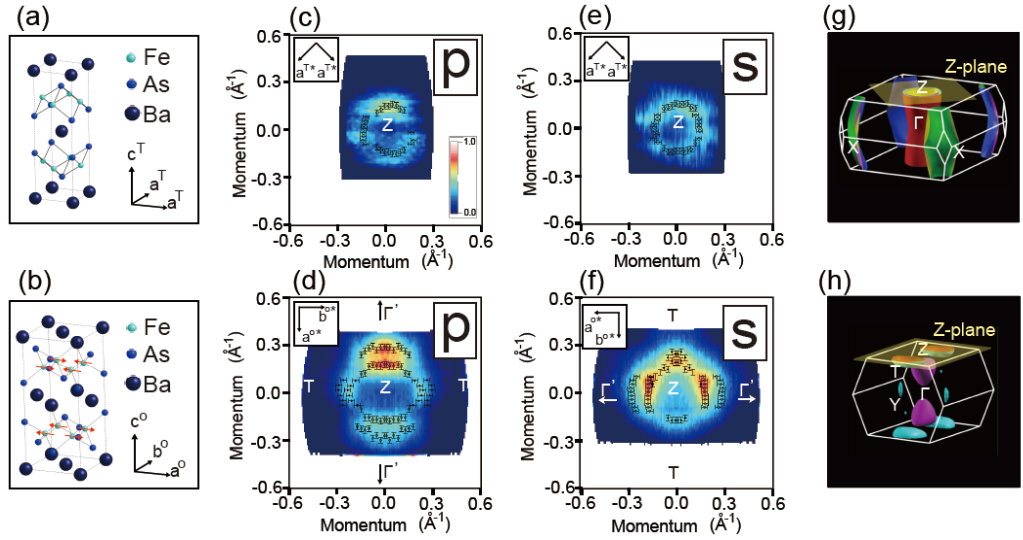

T. Shimojima et al.

Figure 2

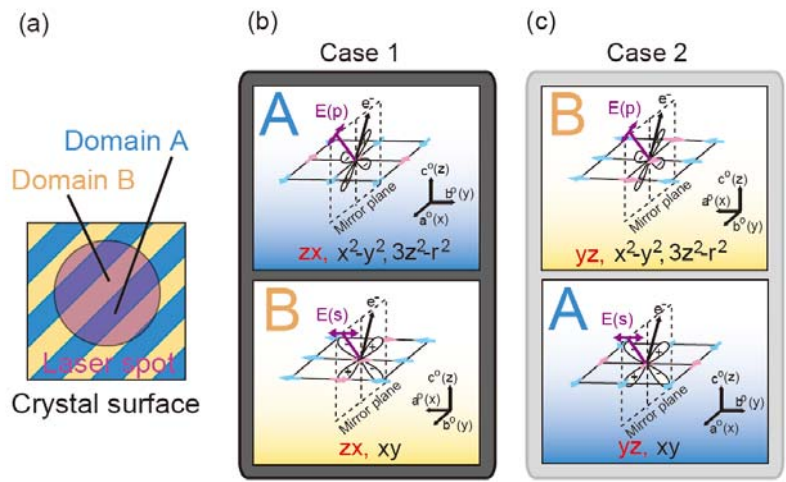

T. Shimojima et al.

Figure 3

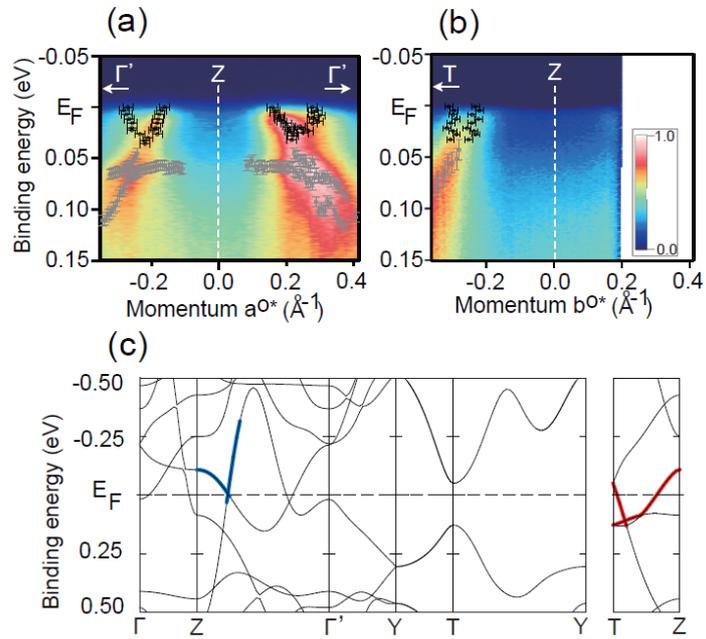

Figure 4

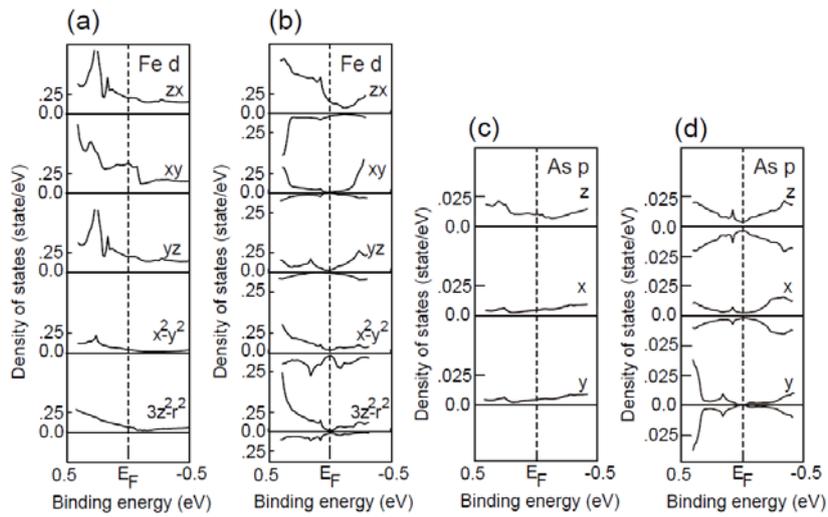

T. Shimojima et al.

# Supplementary figure 1

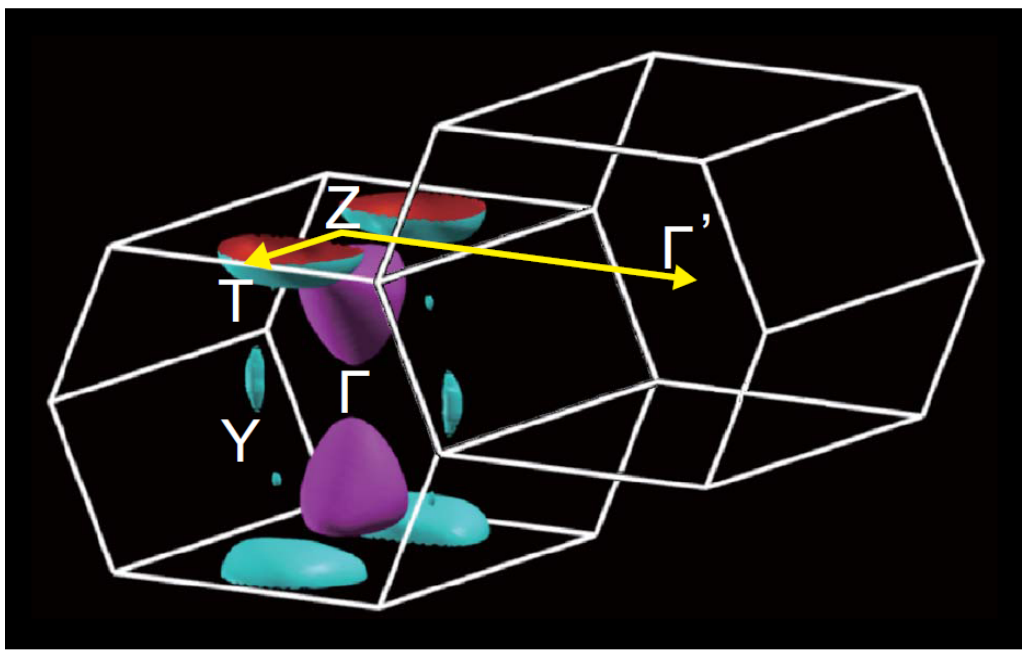

T. Shimojima et al.